\documentclass[aps,prl,twocolumn,floatfix,groupedaddress]{revtex4-1}

\usepackage{amscd,amsfonts,amsmath,amssymb,amsthm}
\usepackage{graphicx}
\usepackage{color}
\usepackage{times}
\usepackage[english]{babel}
\usepackage[pdftex]{hyperref} %pdflatex
\usepackage[latin1]{inputenc}
\newcommand{\abs}[1]{\left\vert #1 \right\vert}

\newcommand{\expect}[1]{\left\langle{#1}\right\rangle}
\newcommand{\He}{\mathrm{He}}

\newcommand{\KL}{Karhunen-Lo\'eve }

\newcommand{\onenorm}[1]{\left\vert\left\vert#1\right\vert\right\vert_1}

\newcommand{\timeorder} {\underset{\leftarrow}{\mathcal{T}}}

\renewcommand{\section}[1]{\vspace{0cm}\noindent\emph{#1} --}
\renewcommand{\subsection}[1]{\vspace{0cm}\noindent\emph{#1} --}
\begin{document}
\title{Simulation of stochastic quantum systems using polynomial chaos expansions}
\author{Kevin C.~Young}
\email[Electronic address: ]{kyoung@sandia.gov}
\author{Matthew D.~Grace}
\email[Electronic address: ]{mgrace@sandia.gov}
\affiliation{Department of Scalable and Secure Systems Research,
  Sandia National Laboratories, Livermore, CA 94550}
\date{\today}

\begin{abstract}
\noindent We present an approach to the simulation of quantum systems
driven by classical stochastic processes that is based on the polynomial
chaos expansion, a well-known technique in the field of uncertainty
quantification. The polynomial chaos expansion represents the
system density matrix as a series of orthogonal polynomials in the 
principle components of the stochastic process and yields a sparsely
coupled hierarchy of linear differential equations. We provide practical
heuristics for truncating this expansion based on results from
time-dependent perturbation theory and demonstrate, via an
experimentally relevant one-qubit numerical example, that our technique
can be significantly more computationally efficient than Monte Carlo
simulation.
\end{abstract}
\maketitle

%-----------------------------------------------------------------
% Section: Introduction % (fold)
%-----------------------------------------------------------------
\section{Introduction} 
\label{sec:introduction}
Quantitative understanding of the dynamics of open quantum
systems is critically important to many contemporary physics experiments
\cite{Breuer02a}. While the equations of motion for open systems models
are often simple to formulate, in only a few special cases may they be
solved analytically, and numerical studies are often limited to small
systems. However, when quantum back action can be neglected,
\emph{i.e.,} at high temperature or for short times, fully quantum open
systems may be well approximated by semi-classical stochastic driving,
whereupon the environment interaction operators are replaced by
classical stochastic processes. For example, the coherence decay of
diamond nitrogen-vacancy (NV) centers in the presence of dilute
paramagnetic defects may be modeled very well by assuming that
paramagnetic defects in the lattice produce a classical, fluctuating
Overhauser field which dephases the NV center \cite{Hanson08a,
 Ladd10a}. Expensive numerical studies modeling the full quantum
environment are then only required only to determine the statistical
properties of this effective field. The resulting stochastic models are
often sufficient to compute any desired system observables. However,
these reduced models exchange quantum degrees of freedom for stochastic
ones that may also require large, but significantly reduced,
computational overhead. Expectation values of the system observables may
then, in principle, be computed by averaging over the stochastic degrees
of freedom in a manner that is consistent with the statistics of the
stochastic process. In practice, however, such an average is often
difficult to calculate. Monte Carlo (MC) methods \cite{Press92a}
approximate this average by generating {many} sample noise trajectories,
integrating the  Schr\"odinger equation for each trajectory, and 
averaging the resulting density operators. However, MC can be
notoriously slow to converge, making it impractical for applications
requiring iterative numerical calculations, such as optimal control
\cite{Grace07a, *Grace07b, Gorman12a}. Perturbative master equations,
on the other hand, are often either computationally inexpensive and
inaccurate, or expensive and accurate, depending on the approximations
made in their derivation \cite{Breuer02a}.

In this work, we present an alternative approach to performing the
stochastic average based on a class of techniques used widely in
classical uncertainty quantification. Known as the polynomial chaos
expansion (PCE) \cite{Debusschere04a, LeMaitre10a}, this method
leverages properties of orthogonal polynomials to yield a converging
sequence of approximate evolution equations for a quantum system
undergoing stochastic driving without resorting to MC methods. While we
restrict our discussion to quantum systems driven by classical Gaussian
stochastic processes, we make no assumptions of weak coupling nor do we
restrict the form of the noise correlation function. Furthermore, we show
that the linearity of the Schr\"odinger equation makes quantum systems
particularly well suited to the PCE approach, as the stochastic dynamics
may be expressed in terms of a sparsely-coupled system of differential
equations.

We begin this article with a derivation of \KL decomposition, which
expresses correlated, classical stochastic processes as an easily
truncated sum of deterministic functions weighted by \emph{uncorrelated}
random variables. We proceed to use this decomposition to derive the
PCE as applied to stochastic quantum systems, yielding a
sparsely-coupled system of Schr\"odinger-like equations. We conclude
with a discussion and numerical simulation of the convergence properties
of this method, benchmarking our results against Monte Carlo
simulations.

%-----------------------------------------------------------------
% Section: Model % (fold)
%-----------------------------------------------------------------
\section{Model}
\label{sec:model}
We consider a quantum two-level system coupled linearly to a classical
stochastic process, $\Omega$, and described by the Hamiltonian,
$H(t;\Omega(t)) = H_0(t) + \Omega(t) V$. Switching to a rotating frame
with respect to $H_0$, we obtain:
\begin{equation}
  \label{eq:hamiltonian}
  \tilde H(t;\Omega(t)) = \Omega(t) U_0(t)^\dagger V U_0 \equiv
  \Omega(t) \tilde V(t),
\end{equation}
where $U_0 = \timeorder \exp(-i\int_0^t H_{0}(s) ds)$ and $\timeorder$
is the Dyson time-ordering operator. Hamiltonians of this form are quite
common, and restriction to this minimal form simplifies the following
derivations. Generalizations to multiple or more complicated dependence
on the stochastic process require only straightforward modifications to
the following procedure.

We restrict our discussion here to stochastic processes, $\Omega$,
which are mean-zero, Gaussian, and stationary \cite{Ripley08}. By
Wick's theorem \cite{Gardiner04b}, such processes may be completely
described in terms of their two-point correlation functions, $C(t_1,
t_2)\equiv \expect{\Omega(t_1)\Omega(t_2)}_{\Omega}$. In this article,
we use the notation $\expect{f(\Omega(t))}_{\Omega}$ to signify the
expectation value of the function $f$ with respect to the
process, $\Omega$.

The state of the system when conditioned on a specific realization of the
stochastic process, $\rho(t;\{\Omega(t)\})$, will evolve according to
the Schr\"odinger--von Neumann equation:
\begin{equation}
  \label{eq:schrod}
  i \frac{d \rho(t;\{\Omega(t)\})}{dt} = \Omega(t) V(t)^\times
  \rho(t;\{\Omega(t)\}), 
\end{equation}
where we have used the superoperator adjoint notation $A^\times B = [A,
  B]$. At a time $\tau$, the state of the system, averaged over the
stochastic process, is given by the formal expression
\begin{equation}
  \label{eq:state_final}
  \rho(\tau) = \expect{U(\tau; \{\Omega(t)\}) \rho(0) U(\tau;
  \{\Omega(t)\})}_{\Omega},
\end{equation}
where $U(\tau;\{\Omega(t)\})$ denotes the unitary operator generated by
Eq.~\eqref{eq:schrod} with a specific realization of $\Omega$. The
objective of this work is to demonstrate that this stochastic average
may be performed in a computationally efficient manner using the
PCE.

\subsection{Karhunen-Lo\'eve Expansion}
If the stochastic process, $\Omega$, is white, the average in
Eq.~\eqref{eq:state_final} may be taken locally in time, and the system
dynamics may be described exactly by a Lindblad master equation
\cite{Breuer02a}. However, the presence of non-vanishing time 
correlations greatly complicates the calculation of the stochastic
average. To simplify this calculation, we employ the \KL expansion
(KLE) \cite{LeMaitre10a}, which expresses a continuous, correlated
process in terms of a discrete sum of deterministic functions weighted
by \emph{uncorrelated} random-variables:
\begin{equation}
 \label{eq:KL}
 \Omega(t) = \sum_{n=1}^\infty \sqrt{\lambda_n} g_n(t) \xi_n. 
\end{equation}
Here, $\xi_n \in\mathcal{N}(0,1)$ are independent and identically
distributed (\emph{iid}) random variables drawn from a unit-variance,
zero-mean Gaussian distribution, while $\lambda_n$ and $g_n(t)$ are,
respectively, the eigenvalues and $L_2$-orthonormal eigenfunctions of
the Fredholm equation \cite{LeMaitre10a}:
\begin{equation}
 \label{eq:fredholm}
 \int_{0}^\tau C(t_1,t_2) g_n(t_2) dt_2= \lambda_n g_n(t_1).
\end{equation}
Here, the correlation function acts as a symmetric, positive
semi-definite integral kernel, so Mercer's theorem \cite{LeMaitre10a}
implies that the eigenvalues, $\lambda_n$, are discrete and
non-negative. Non-negativity is ensured because the correlation
functions of stationary processes are positive semi-definite (by
Bochner's theorem \cite{Reed80a}), while discreteness is guaranteed by
the finite upper limit on the integral, Eq.~\eqref{eq:fredholm}.

\begin{figure}
  \begin{center}
    \includegraphics[width=\columnwidth]{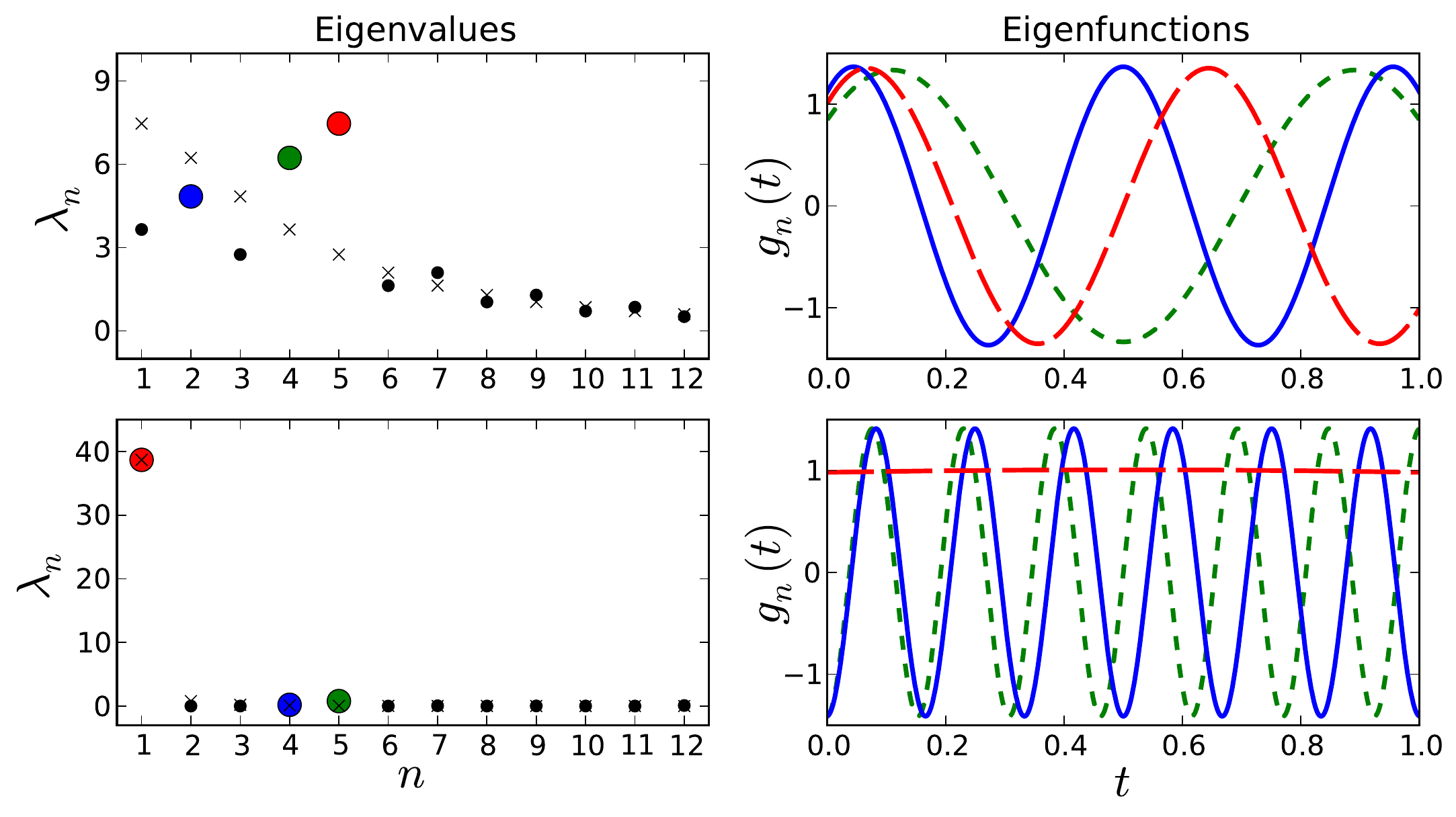}
    \caption{(color online) The eigenvalues, $\lambda_n$ (left, gray
      $\times$'s); cumulative transition rates, $\Gamma_n$,
      (left, solid circles); and eigenfunctions, $g_n(t)$ (right), of
      the Fredholm equation, Eq.~\eqref{eq:fredholm}, for
      Orenstein-Uhlenbeck noise, where $C(t) = \exp(-\abs{t}/\tau_c)$,
      calculated with final time $\tau = 1$ and Hamiltonian $H(t) =
      B\sigma_x + \Omega(t)\sigma_z$. The top row corresponds to noise
      correlation time $\tau_c = 0.1$ and magnetic field $B = 5$, while
      the bottom row for $\tau_c = 10$ and magnetic field $B = 20$. The
      stochastic modes with the three largest transition rates are color
      coordinated with the eigenfunction plots (in descending transition
      rates: red dot and long dashed line; green dot and short dashed
      line; blue dot and solid blue line). For long correlation times,
      compared to $\tau$, only one mode is dominant, while more modes
      are increasingly important for shorter correlation times.} 
    \label{fig:kl}
  \end{center}
\end{figure}

Interestingly, when the final time is much greater than the correlation
time of the stochastic process, \emph{i.e.,} $\tau \gg \tau_c$, the 
Wiener-Khinchin theorem \cite{Gardiner04b} implies that the eigenvalue
spectrum becomes continuous and equal to the noise power spectral
density, \emph{i.e.,} $\lambda_\omega = S(\omega)$, while the
eigenfunctions take the form $g_\omega(t) \propto \cos(\omega t)$. Taken
to the extreme white-noise limit, where $\tau_c \rightarrow 0$,
\emph{all} eigenvalues are equal. In the regime where $\tau \rightarrow
\tau_c$, the correlation function is approximately constant over the
integration window, and the Fredholm equation possesses only a single
nonzero eigenvalue, with eigenfunction $g_1(t) \propto 1$. In this case,
the stochastic process may be well approximated as a random variable
which is constant over $t \in [0,\tau]$. Figure~\ref{fig:kl} illustrates
these two limits by plotting the eigensystem of the Fredholm equation
for Orenstein-Uhlenbeck \cite{Gardiner04b} noise with two different
decay parameters.

When the correlation time is relatively long compared to the evolution
time, but not infinite, \emph{e.g.,} when $\tau < \tau_c$, the expansion
Eq.~\eqref{eq:KL} is dominated by only a few terms corresponding to the
largest $S$ eigenvalues, and so may be truncated with minimal error.
However, truncating the KLE based only on the eigenvalues ignores the
impact that higher frequency modes could have on the dynamics,
\emph{e.g.,}, resonance. We therefore propose a more physically motivated
truncation criterion based on results from time-dependent perturbation
theory \cite{Sakurai94a}. For static Hamiltonian terms $H_0$ and $V$ in
the Schr\"odinger picture, the rate at which any given mode, $g_n(t)$,
could cause a transition between the $j$th and $k$th eigenstates of
$H_0$ is given by
\begin{equation}
  \label{eq:rate}
  \Gamma_n^{jk} = \frac{1}{\tau} 
  \abs{\left\langle j \abs{V} k \right\rangle 
  \int_0^\tau e^{i(E_j - E_k)t} \sqrt{\lambda_n} g_n(t) dt}^2,
\end{equation}
where $H_{0} |j\rangle = E_{j} |j\rangle$. Summing over these
eigenstates provides a measure of the degree to which a given mode will
impact the evolution of the system, \emph{i.e.,} the cumulative
transition rate, $\Gamma_n = \sum_{j,k} \Gamma_n^{jk}$. In addition to
the eigenvalues, $\lambda_n$, and eigenfunctions, $g_n(t)$, of the
Fredholm equation, Eq.~\eqref{eq:fredholm}, cumulative transition rates
are also included in Figure~\ref{fig:kl}. Thus, we approximate the
expansion Eq.~\eqref{eq:KL} by keeping only those modes corresponding to 
the $S$ largest transition rates, $\Gamma_n$; we shall refer to $S$ as
the \emph{stochastic dimension}. With this approximation,
Eq.~\eqref{eq:schrod} becomes 
\begin{equation}
 \label{eq:noomega}
 i\frac{d \rho(t;\vec \xi)}{dt} 
 = \sum_{n=1}^S \sqrt{\lambda_n} g_n(t) \xi_n V(t)^\times \rho(t;\vec\xi).
\end{equation}
We emphasize that this truncation strategy differs from that usually
taken in standard uncertainty quantification literature
\cite{LeMaitre10a}, wherein the truncation is based on the magnitude of 
the eigenvalues alone, and without consideration to the potential impact
of a given stochastic mode on the system dynamics.

%-----------------------------------------------------------------
% Subsection: Expansion in orthogonal polynomials % (fold)
%-----------------------------------------------------------------
\subsection{Expansion in orthogonal polynomials}
\label{sub:expansion_in_orthogonal_polynomials}
At the final time, $\tau$, the state of the system may be considered as
a complicated function of the $S$ uncorrelated random variables from
Eq.~\eqref{eq:KL}, \emph{i.e.,} $\rho(\tau) = \rho(\tau;\vec\xi)$, as
expressed in Eq.~\eqref{eq:noomega}. As such, we may expand this
function in a complete basis of orthogonal polynomials,
$\Phi_n(\vec\xi)$:
\begin{equation}
  \label{eq:pce1}
  \rho(t;\vec\xi) = \sum_{n=0}^\infty \phi_n(t) \Phi_{n}(\vec\xi\,),
\end{equation} 
yielding the (untruncated) PCE. Here, $\phi_n(t)$ are the
time-dependent, operator-valued expansion coefficients that represent
our new dynamical variables. The polynomials should be orthogonal under
the measure, $\mu d\xi = \exp(-\xi^2/2)/\sqrt{2\pi} d\xi$, which is
derived from the stationary distribution of the random variables, which
are drawn from $\mathcal{N}(0,1)$. Multivariate Hermite polynomials are
the natural choice:
\begin{equation*}
  \Phi_\mathbf{n}(\vec \xi\,) = \prod_{j=1}^S \He_{{n}_j}(\xi_j),
\end{equation*}
where we now use the multi-index vector $\mathbf{n}
\in\mathbb{Z}_{\geqslant0}^{S}$. This expansion, Eq.~\eqref{eq:pce1},
may be truncated, keeping only terms for which $\onenorm{\mathbf{n}} =
\sum_j n_j \le P$, where $P$ is the PCE \emph{order}. Note that with
this truncation the PCE, like general second-order master equations
\cite{Breuer02a}, is no longer guaranteed to preserve the positivity of
the density matrix.  However, in practice, we have seen no violations of positivity, and if negative eigenvalues were to appear, they could likely be eliminated by moving to a higher-order expansion.
Inserting the truncated expansion into the
evolution equation, Eq.~\eqref{eq:noomega}, we have
\begin{equation*}\!\!
  \sum_{\onenorm{\mathbf{k}}=0}^P \!\!i \frac{d \phi_\mathbf{k}(t)}{dt}
  \Phi_\mathbf{k}(\vec\xi\,)
  = \sum_{n=1}^S \sqrt{\lambda_n} g_n(t) \xi_n V(t)^\times \!\!\!\!
  \sum_{\onenorm{\mathbf{l}}=0}^P \!\!\!\! \phi_\mathbf{l}(t)
  \Phi_\mathbf{l} (\vec\xi\,).
\end{equation*}
Exploiting the orthogonality of the Hermite polynomials, we may compute
the evolution equation for the coefficients of the expansion of
Eq.~\eqref{eq:pce1}: 
\begin{equation}
 \label{eq:expansion}
 i\frac{d\phi_\mathbf{m}(t)}{dt} = \sum_{n=1}^S \sum_{\onenorm{\mathbf{l}}=0}^P
 \sqrt{\lambda_n} g_n(t) V(t)^\times \phi_\mathbf{l}(t)
 G_{\mathbf{m}n\mathbf{l}},
\end{equation}
where we have defined the \emph{Galerkin projection}:
\begin{equation}
 \label{eq:galerkin}
 G_{\mathbf{m}n\mathbf{l}} = \frac{\expect{\Phi_\mathbf{m}(\vec\xi\,)
  \xi_n \Phi_\mathbf{l}(\vec\xi\,) }_\xi} 
 {\expect{\Phi_\mathbf{m}(\vec\xi)^2}_\xi}.
\end{equation}
These projection terms may be computed explicitly using the Hermite
polynomial orthogonality relations:
\begin{equation*}
 \int_{-\infty}^{\infty} \He_n(x)\He_m(x)e^{-x^2/2} dx = \sqrt{2\pi}
 n! \delta_{m,n},
\end{equation*}
and the three term recurrence relation:
\begin{equation*}
 \xi \He_n(\xi) = \He_{n+1}(\xi) + n \He_{n-1}(\xi).
\end{equation*}
Taken together, these lead to a much-simplified expression for the
Galerkin projection:
\begin{align*}
 G_{\mathbf{m}n\mathbf{l}}
 &= \left((\mathbf{m}_n+1)\delta_{\mathbf{m}_n+1,\mathbf{l}_n} 
 + \delta_{\mathbf{m}_n-1,\mathbf{l}_n}\right)\prod_{j\ne n}
 \delta_{\mathbf{m}_j \mathbf{l}_j}
\end{align*}

Owing to the presence of the Kronecker delta functions in the Galerkin
projection, the PCE results in a sparsely-coupled hierarchy of 
deterministic linear differential equations which can be solved by 
standard numerical methods.  The choice of both the stochastic dimension
{$S$} and the PCE order {$P$} determine the number of
equations {$N$} in the hierarchy through a simple combinatorial
argument \cite{LeMaitre10a}:
\begin{equation}
  \label{eq:factorial}
  N = \sum_{m = 0}^P {{S+m-1}\choose{S-1}} = \frac{(S+P)!}{S!\,P!}.
\end{equation}
This scaling, known colloquially as the \emph{curse of dimensionality},
limits practical applications to those situations in which i) the noise
correlation time is long, resulting in low stochastic dimension and 
ii) the noise is weak, so that the PCE converges quickly. More
sophisticated truncation procedures may reduce the hierarchy depth,
however, this remains an area of active research.

%-----------------------------------------------------------------
% Section: Convergence of the PCE % (fold)
%-----------------------------------------------------------------
\section{Convergence of the PCE} 
\label{sec:convergence_of_the_pce}
The convergence properties of the coupled evolution equations,
Eq.~\eqref{eq:expansion}, depend critically on the distribution of the
cumulative transition rates, $\Gamma_n$. Specifically, noise modes
associated with large transition rates will couple strongly to the
system and the PCE must be truncated at high order in those variables to
faithfully represent the system dynamics. For example, modes for which
$\Gamma_n \tau > 1$ will, on average, induce at least one transition
over the course of the evolution. Accurately capturing these dynamics
would require such modes to be considered at high PCE order.

\begin{figure}
  \begin{center}
    \includegraphics[width=\columnwidth]{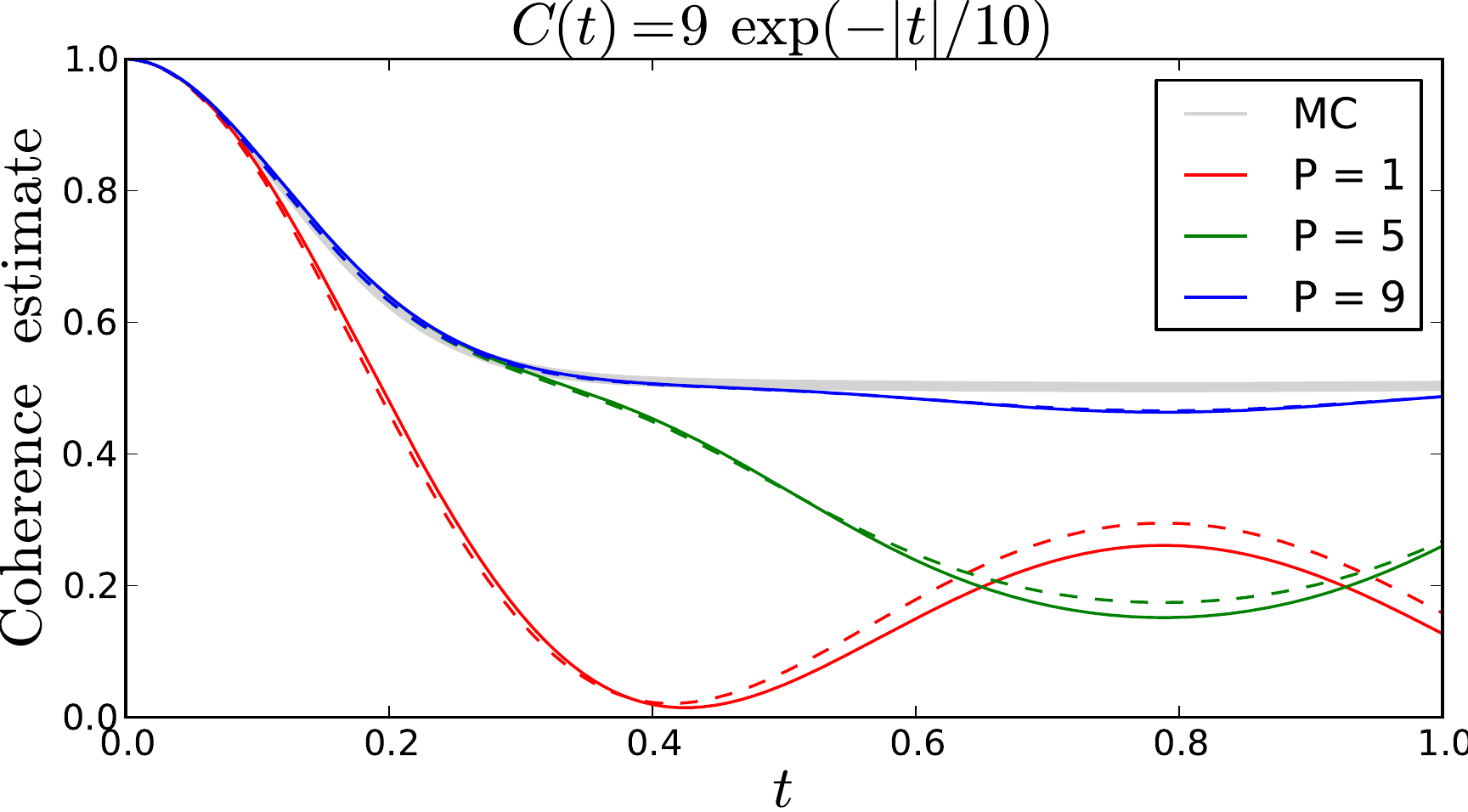}
    \caption{(Color online) A comparison of the predicted time-dependent
      coherence, $\expect{\sigma_x(t)}$, for the Hamiltonian given in
      Eq.~\eqref{eq:model}, with correlation function $C(t) = 9
      \exp(-\abs{t}/10)$. Solid, colored lines correspond to a
      stochastic dimension $S=1$, while dashed lines correspond to
      $S=3$. Grey line is the Monte Carlo result drawn with width equal to
      the standard error in the estimate. Because of the long
      correlation time, only one eigenvalue of the KLE is dominant, so
      it is more efficient to keep the stochastic dimension small and
      increase the PCE order.}
    \label{fig:compare}
  \end{center}
\end{figure}
 
We consider explicitly the stochastic dynamics of a driven quantum
two-level system, or qubit, coupled to a classically fluctuating
dephasing process:
\begin{equation}
 \label{eq:model}
 H(t) = \sigma_x + \Omega(t) \sigma_z,
\end{equation}
where $\sigma_x$ and $\sigma_z$ are Pauli matrices. Such a model
describes, for example, Rabi oscillations in the presence of dephasing
noise \cite{Gorman12a, Young12a}, and is particularly relevant for NV
centers in diamond \cite{Hanson08a, Ladd10a}. Other examples of relevant
stochastically-driven systems include dephasing noise in trapped ions
\cite{Biercuk09a, Uys09a} and $1/f$ noise in superconducting qubits
\cite{Clarke08a}. In the absence of the drift term $\sigma_{x}$, the
dephasing dynamics are exactly solvable for any stationary Gaussian
process, $\Omega$. However, when this term is included, the Hamiltonian
does not commute with itself at different times, and the system is no
longer analytically integrable. To illustrate our method, we choose the
stochastic process, $\Omega$, to be Gaussian Orenstein-Uhlenbeck type
\cite{Gardiner04a}, with correlation function of the form $C(t) =
\alpha^2 \exp(-\abs{t}/\tau_c)$; the coupling parameter, $\alpha$, and
the correlation time, $\tau_c$, will be specified later. We specify the
initial state of the system as $|\sigma_{x}^{+}\rangle$, where
$\sigma_{x} |\sigma_{x}^{\pm}\rangle = \pm |\sigma_{x}^{\pm}\rangle$,
and we compute the time-dependent coherence, $\expect{\sigma_x(t)}$. By
tuning the noise correlation time and the coupling parameter, this model
can explore the convergence of our method with respect to PCE order $P$
and stochastic dimension $S$.

To benchmark the performance of our PCE approach, we compare against MC
simulations. MC algorithms approach the problem of computing the 
stochastic average in Eq.~\eqref{eq:state_final} by generating a
sufficiently large number of statistically consistent realizations of
the stochastic process, $\Omega$, evolving the system with each of the
realizations, and averaging the final-time density matrices.

As shown in Fig.~\ref{fig:compare}, our PCE method is capable of
reproducing the results of MC simulations with high accuracy,
significantly faster than MC. For the example chosen, Monte Carlo
required approximately 4000 iterations for convergence, while the most
accurate PCE results report here ($P = 9$ and $S = 3$) required the
solution of only 220 coupled equations and ran approximately 20 times
faster than the MC simulation. Note that because $\tau_{c}/\tau = 10$ in
our simulation, only one eigenvalue of the KLE is dominant. In this
regime, it is more efficient to keep the stochastic dimension small ($S
\leq 3$) and increase the PCE order for improved accuracy. As the order
increases from $P = 1$ to $P = 9$, the accuracy of the PCE coherence
increases as a function of time, compared to the converged MC result.

%-----------------------------------------------------------------
% Section: Conclusion % (fold)
%-----------------------------------------------------------------
\section{Discussion} 
\label{sec:conclusion}
Our PCE method demonstrates the ability to rapidly and accurately
propagate stochastic quantum systems. It outperforms MC simulations in
computational efficiency, and has the potential to become an important
tool in the study of noisy quantum systems. An area in which we expect
the PCE to be particularly useful is in the realm of optimal control
(OC). The high computational cost of MC simulations severely limits its
use in sequential optimal control simulations. However, PCEs can be both
accurate and fast, and they may be easily incorporated as part of a
surrogate dynamical model in OC simulations. In this work, the PCE has
been formulated to propagate a particular state, however, the equation
of motion for the complete dynamical map takes a similar form and may
also be adapted easily to PCE methods. Implementation of state-to-state
and dynamical map OC will appear in forthcoming work. 

An interesting comparison can be made between the PCEs as presented here 
and another expansion based on orthogonal polynomials: Kubo's hierarchy
equations of motion (HEOM) \cite{Kubo69a}, and their generalization to
all diffusive processes, the DHEOM \cite{Sarovar12a}. Application of the
DHEOM/HEOM demands that the noise be diffusive and have exponentially
decaying correlation function, and proceeds by diagonalizing the noise
generating functional using of orthogonal polynomials
\cite{Sarovar12a}. Interestingly, though the HEOM and PCE approaches 
each yields a sparsely-coupled hierarchy of differential equations based
on expansions in orthogonal polynomials, they perform well in exactly
opposite limits: the HEOM method converges quickly for noise with a
short correlation time, while the stochastic dimension of our PCE method
converges quickly for noise with a long correlation time. For systems
coupled to multiple, uncorrelated noise sources, it may be
computationally advantageous to apply different methods for each source:
HEOM for noise with short correlation times, PCEs for noise with long
correlation times.

Additionally, we have assumed a linear coupling between the system and
the stochastic process in Eq.~\eqref{eq:hamiltonian}. While such a
coupling is common \cite{Joynt10}, nonlinear interactions are possible,
which may increase the coupling density of the differential equations in
Eq.~\eqref{eq:pce1} by modifying the form of the Galerkin projection of
Eq.~\eqref{eq:galerkin}. Strongly nonlinear interactions will yield
densely coupled systems of equations, which will increase the
computational cost of this method.

Despite the obvious utility of our PCE approach for simulating
stochastic quantum systems, it does have limitations. Principal among
these is the uncontrolled approximation of the stochastic average, so
that the error must be estimated by increasing the PCE order and/or the
stochastic dimension of the expansion until convergence is
seen. Furthermore, as indicated in Eq.~\eqref{eq:factorial}, the number
of equations to be solved grows combinatorially with both the stochastic
dimension and the PCE order; when these are large, the method of
Galerkin projections becomes computationally infeasibile. Intermediate
between MC and the PCE approach presented here is a \emph{non-intrusive}
formulation of the PCE, so called because its implementation requires
only a forward solver for the equations of motion, while the
\emph{intrusive} method presented here requires an explicit solver to
propagate the coupled equations resulting from the Galerkin
projections. Non-intrusive spectral methods approximate the stochastic
average of Eq.~\eqref{eq:state_final} by performing a KLE, and using
sparse quadrature techniques to perform the average. We plan to
implement such techniques in the near future.

%-----------------------------------------------------------------
% Section: Acknowledgements % (fold)
%-----------------------------------------------------------------
\section{Acknowledgements} 
\label{sec:acknowledgements}
We gratefully acknowledge enlightening discussions with Constantin Brif,
Cosmin Safta, Maher Salloum, and Mohan Sarovar (SNL-CA). This work was
supported by the Laboratory Directed Research and Development program at
Sandia National Laboratories. Sandia is a multi-program laboratory
managed and operated by Sandia Corporation, a wholly owned subsidiary of
Lockheed Martin Corporation, for the United States Department of
Energy's National Nuclear Security Administration under contract
DE-AC04-94AL85000.

\bibliography{./pce}
\bibliographystyle{apsrev4-1}

\end{document}